\documentclass[journal,10pt]{IEEEtran}
\usepackage{geometry}
\usepackage{diagbox}
\usepackage{multirow}
\usepackage{graphicx}
\usepackage[fleqn]{amsmath}
\usepackage{stfloats}
\usepackage{cite}
\usepackage{algorithm}
\usepackage{algorithmic}

\usepackage{amssymb}
\usepackage{color}

\begin{document}
\title{Intermittent Encryption Strategies for Anti-Eavesdropping Estimation}
\author{Zhongyao~Hu, Bo~Chen,~\IEEEmembership{Senior Member,~IEEE}, Pindi Weng, Jianzheng Wang, Li~Yu,~\IEEEmembership{Senior Member,~IEEE}
\thanks{Manuscript received xx; accepted xx. Date of publication xx; date of current version xx. This work was supported in part by the National Natural Science Funds of China under Grant 61973277 and Grant 62073292, in part by the Zhejiang Provincial Natural Science Foundation of China under Grant LR20F030004, in part by the Key Research and Development Program of Zhejiang Province under Grant 2022C03029 and Grant
2023C01144. This brief was recommended by Associate Editor xx. {\em (Corresponding author: Bo Chen.)}}
\thanks{Z. Hu, B. Chen, P. Weng and L. Yu are with Department of Automation, Zhejiang University of Technology, Hangzhou 310023, China. (email: huzhongyao@aliyun.com, bchen@aliyun.com, pwd2gg@aliyun.com, lyu@zjut.edu.cn).}
\thanks{J. Wang is with the School of Electrical and Electronic Engineering, Nanyang Technological University, Singapore 639798. (email: wang1151@e.ntu.edu.sg).}
}
\markboth{Preprint, 2024}
{Shell \MakeLowercase{\textit{et al.}}: Bare Demo of IEEEtran.cls for Journals}
\maketitle

\begin{abstract}
In this paper, an anti-eavesdropping estimation problem is investigated. A linear encryption scheme is utilized, which first linearly transforms innovation via an encryption matrix and then encrypts some components of the transformed innovation.
To reduce the computation and energy resources consumed by the linear encryption scheme, both stochastic and deterministic intermittent strategies which perform the linear encryption scheme only at partial moments are developed. When the system is stable, it is shown that the mean squared error (MSE) of the eavesdropper converges under any stochastic or deterministic intermittent strategy. Also, an analytical encryption matrix that maximizes the steady state of the MSE is designed. When the system is unstable, the eavesdropper's MSE can be unbounded with arbitrary positive encryption probabilities and decision functions if encryption matrices are chosen appropriately. Then, the relationship between the aforementioned encryption parameters and the eavesdropper's MSE is analyzed. Moreover, a single intermittent strategy which only encrypts one message is discussed. This strategy can be unavailable for stable systems, but can make the eavesdropper's MSE unbounded in unstable systems for the encrypted message satisfies a linear matrix inequality (LMI) condition. The effectiveness of the proposed methods is verified in the simulation.
\end{abstract}


\vspace{-3pt}
\section{Introduction}
The estimation of system states with sensor data is of great importance for many engineering problems such as control and detection \cite{Optimalfiltering}. Nowadays, wireless technologies are often used to transmit sensor data because of their advantages in lightweight and efficiency. However, due to the broadcast nature of wireless media, illegal eavesdroppers in the vicinity can overhear the content in the data channels. Such a data leakage can compromise the confidentiality of users and may also be used to perform more lethal attacks \cite{TEIXEIRA2015135}. Therefore, it is particularly important to prevent eavesdropping. Currently, injecting artificial noise into original data is one of the most common defense mechanisms against eavesdropping. In this mechanism, the artificial noise needs to be designed appropriately to ensure that both the user and the eavesdropper can satisfy some performance constraints \cite{Leong8550317,Mo7465717,Le6606817,Xu10054451}. Artificial packet loss is another common defense mechanism, which can reduce the estimation performance of the eavesdropper by actively retaining sensor data \cite{TSIAMIS20178385,Impicciatore9992668,Leong8543618}. However, this mechanism may also make the mean squared error (MSE) of the user larger.
Recently, a secrete code mechanism was proposed in \cite{Tsiamis8758381}, which does not transmit the original data but a linear combination of multiple estimates. It has been shown in \cite{Tsiamis8758381} and \cite{9762536} that the eavesdropper can only obtain biased data once the eavesdropper loses a packet received by the user, where the bias will be continuously amplified over time. Nevertheless, employing this mechanism requires configuring an additional feedback channel, which may increase undesired communication burdens.
Compared with the mechanisms above, encrypting data packets can protect privacy without feedback loops or performance loss, which reveals a considerable potential for application in control and estimation communities. A privacy-preserving cloud control system is designed in \cite{Ali10269762} by developing a low-complexity chaotic encryption algorithm.. In \cite{Zhang10241989}, the chaotic encryption was used to mask the pixels of images to achieve privacy-preserving perception-driven control. \cite{Suryavanshi10185646} proposed a closed-loop encrypted model predictive controller using Paillier Cryptosystem. Moreover, based on homomorphic encryption techniques, \cite{LU2018314} and \cite{Hadjicostis8967106} investigated privacy problems in distributed optimization and average consensus, respectively.

It should be emphasized that the encryption-based methods tend to be computationally intensive due to the complexity of the encryption algorithms \cite{ASHIBANI201781}. Therefore, for wireless sensors with limited energy and computation resources, it is crucial to reduce the amount of data being encrypted. \cite{YANG2020109116} proposed a privacy-preserving estimator based on a fast stochastic encryption mechanism, but the resulting transmission consumption is high due to the high dimension of the ciphertext \cite{Vilajosana6627960}. The works in \cite{HUANG2021109537} and \cite{WANG2022110145} allow the sensor to switch between encryption and non-encryption modalities, and design scheduling algorithms to balance the encryption consumption and the privacy level. However, the algorithms in \cite{HUANG2021109537} and \cite{WANG2022110145} may not be usable in the systems with high state dimensions because encrypting the entire state in those cases is bulky. Moreover, the authors of \cite{Shang9882330} proposed a linear encryption scheme that only encrypts partial components of the transformed data, thus overcoming the drawbacks of the methods in \cite{HUANG2021109537} and \cite{WANG2022110145}. Nevertheless, the method in \cite{Shang9882330} involves high energy consumption since the linear encryption scheme is performed at every moment. To overcome the aforementioned drawbacks, this paper will investigate intermittent encryption strategies, i.e., the encryption will be performed only at partial moments. Specifically, three kinds of intermittent strategies will be considered: 1) stochastic intermittent strategy that determines whether to encrypt or not based on a probability, 2) deterministic intermittent strategy that determines whether to encrypt or not based on a periodic decision function, and 3) single intermittent strategy that encrypts only one packet. Based on the discussions above, the contributions of this paper are summarized as follows.
\begin{itemize}
\item In stable systems, the MSE of the eavesdropper is shown to converge under any stochastic or deterministic intermittent strategy. The analytical expression of encryption parameters that maximizes the steady state of the MSE is derived. It is proved that the single intermittent strategy is unable to degrade the steady state of the MSE of the eavesdropper.
\item In unstable systems, we show that arbitrary positive encryption probability or period decision function will make the MSE of the eavesdropper unbounded if the encryption parameters are suitably chosen. An analytical expression of the encryption parameters is given. Then, a linear matrix inequality (LMI) condition is provided to determine the encrypted packet in the single intermittent strategy such that the MSE of the eavesdropper is unbounded.
\end{itemize}
\textbf{Notations:} ${\mathbb{R}}^r$ and ${\mathbb{R}}^{r\times s}$ denote the $r$ dimensional and $r\times s$ dimensional Euclidean spaces, respectively. $\mathbb{N}_+$ denotes the positive integer set. The notations $[\cdots,\cdots,\cdots]$ and $[\cdots;\cdots;\cdots]$ indicate horizontal and vertical concatenations, respectively. $\mathrm{Re}(\cdot)$ and $\mathrm{Im}(\cdot)$ represent real and imaginary parts of a complex matrix, respectively. $I$ stands for identity matrix. $\mathrm{Tr}(\cdot)$, $\mathrm{Det}(\cdot)$, $\mathrm{Ker}(\cdot)$, $\mathrm{R}(\cdot)$, $\mathrm{Orth}(\cdot)$ and $\rho(\cdot)$ represent the trace, determinant, null space, range, orthonormal basis for the range and spectral radius of a matrix, respectively. For $X,Y\in{\mathbb{R}}^{r\times r}$, $X>Y$ ($X\geq Y$) means that $X-Y$ is positive definite (positive semi-definite). $\mathbb{E}[\cdot]$ denotes mathematical expectation. $\mathrm{Pr}(\cdot)$ denotes the probability of a random event. $\mathrm{Cov}(\cdot)$ and $\mathrm{Cov}(\cdot,\cdot)$ denote covariance and cross-covariance, respectively. For random variables $x_i\in \mathbb{R}^{r_i}$, $i=1,\cdots,n$, the linear manifold generated by $x_1,\cdots,x_n$ is defined as a set of random variable $x\in \mathbb{R}^r$, that is,  $\mathcal{L}(x_1,\cdots,x_n)\triangleq\{x|x=\sum^n_{i=1}w_ix_i+c,\forall w_i\in\mathbb{R}^{r\times r_i},i=1,\cdots,n,\forall c\in\mathbb{R}^r\}$.
\section{Preliminary}
\subsection{System Description}
Consider the following state space model:
\begin{equation}
\left\{ \begin{array}{l}
x_k=Ax_{k-1}+Bw_{k-1},\\
z_k=Cx_k+v_k,
\end{array} \right.
\label{eq.1}
\end{equation}
where $x_k\in \mathbb{R}^n$ is the system state and $z_k\in \mathbb{R}^m$ is the measurement, $w_k$ and $v_k$ are uncorrelated white Gaussian noises with covariance $Q>0$ and $R>0$. The initial state $x_0$ is Gaussian with mean $x_{0|0}$ and covariance $P_{0|0}\geq0$, which is uncorrelated with the system noises. $C$ is full row rank, $(A,B)$ is stabilizable and $(C,A)$ is detectable. In particular, $C$ being full row rank is without loss of generality, since we can compress a rank-deficiency $C$ losslessly into a full-row-rank matrix \cite{HU2024111520}.

In the smart sensor, the following Kalman filter is employed:
\begin{equation}\begin{aligned}
x^s_{k|k-1}=Ax^s_{k-1|k-1},
\label{eq.2}
\end{aligned}\end{equation}
\begin{equation}\begin{aligned}
P^s_{k|k-1}=AP^s_{k-1|k-1}A^T+BQB^T,
\label{eq.3}
\end{aligned}\end{equation}
\begin{equation}\begin{aligned}
K^s_k=P^s_{k|k-1}C^T(CP^s_{k|k-1}C^T+R)^{-1},
\label{eq.4}
\end{aligned}\end{equation}
\begin{equation}\begin{aligned}
x^s_{k|k}=x^s_{k|k-1}+K^s_k(z_{k}-Cx^s_{k|k-1}),
\label{eq.5}
\end{aligned}\end{equation}
\begin{equation}\begin{aligned}
P^s_{k|k}=&(I-K^s_kC)P^s_{k|k-1},
\label{eq.6}
\end{aligned}\end{equation}
where $x^s_{0|0}=x_{0|0}$, $P^s_{0|0}=P_{0|0}$, $x^s_{i|j}\triangleq \mathbb{E}[x_i|z_1,\cdots,z_j]$ and $P^s_{i|j}\triangleq \mathrm{Cov}(x_i-\mathbb{E}[x_i|z_1,\cdots,z_j])$. At each sampling moment, the smart sensor transmits the innovation $\varepsilon_k\triangleq z_k-Cx^s_{k|k-1}$ to the legitimate user. To prevent eavesdropping, the smart sensor will encrypt $\varepsilon_k$ before transmission. The specific encryption strategy will be described in the next section.
%

\subsection{Linear Encryption Scheme}
In this paper, a linear encryption scheme is employed. First, left multiplying $\varepsilon_k$ by a non-singular matrix $\tilde{S}=[\bar{S}^T,S^T]^T\in \mathbb{R}^{m\times m}$ to obtain $\tilde{S}\varepsilon_k$, where $\bar{S}\in \mathbb{R}^{\bar{m}\times m}$, $S\in \mathbb{R}^{(m-\bar{m})\times m}$ and $1\leq\bar{m}\leq m$. Then, the encryption process can be given by
\begin{equation}\begin{aligned}
\xi_k=\mathcal{E}(\bar{S}\varepsilon_k,\kappa),
\end{aligned}\end{equation}
where $\xi_k$ is the ciphertext, $\kappa\in\mathbb{R}^1$ is the key and $\mathcal{E}:\mathbb{R}^{\bar{m}}\times\mathbb{R}\rightarrow\mathbb{R}^{\bar{m}}$ is the encryption function \cite{Shang9882330}. Then, the sensor will transmit $\xi_k$ and $S\varepsilon_k$. Notice that $\bar{m}=m$ implies that the entire innovation $\varepsilon_k$ is encrypted, in which case $\tilde{S}=\bar{S}$ and $S=\emptyset$. To follow up with a unified framework for analysis, in the rest of this paper, we will use the notation $S=0$ instead of $S=\emptyset$ when $\bar{m}=m$.

In the proposed framework, $A$, $B$, $C$, $Q$, $R$, $\tilde{S}$, $x_{0|0}$ and $P_{0|0}$ are public information known to the user and the eavesdropper. The key $\kappa$ is private information known only to the user. After receiving the messages $\xi_k$ and $S\varepsilon_k$, the user can recover $\varepsilon_k$ and the Kalman filter (2)-(6) using the key $\kappa$ and the non-singularity of $\tilde{S}$. Although $\xi_k$ and $S\varepsilon_k$ are also overheard by the eavesdropper, the ciphertext $\xi_k$ cannot be decrypted due to the absence of the key $\kappa$. As a result, the eavesdropper only knows $S\varepsilon_k$ and cannot recover $\varepsilon_k$ and the Kalman filter (2)-(6). With the discussion above, successively encrypting $\varepsilon_k$ with $\bar{m}=m$ for all $k\in\mathbb{N}_+$ is one of the feasible methods to prevent eavesdropping. However, successive encryption is unaffordable for the wireless sensor with limited energy and computation resources if the computational burden of encryption algorithms is quite high \cite{Cao7377010,ASHIBANI201781}. To address this issue, some intermittent strategies are introduced in the next section.
\subsection{Intermittent Strategies}
Fig. 1 illustrates the system framework under the intermittent strategy, where the binary variable  $\varsigma_k=1$ means that (7) is performed and $\varsigma_k=0$ otherwise.
Accordingly, the information set available to the eavesdropper at $k$th moment is given by
\begin{equation}\begin{aligned}
\mathcal{I}_{1:k}=\{\mathcal{I}_1,\cdots,\mathcal{I}_k\},
\end{aligned}\end{equation}
where $\mathcal{I}_k=\{\varsigma_k,y_k\}$ and
\begin{equation}\begin{aligned}
y_k=
\left\{ \begin{array}{l}
\varepsilon_k,\ \varsigma_k=0,\\
S\varepsilon_k,\ \varsigma_k=1.
\end{array} \right.
\end{aligned}\end{equation}
With (8), the minimum MSE (MMSE) estimate and covariance of the eavesdropper are expressed as $x_{k|k}=\mathbb{E}[x_k|\mathcal{I}_{1:k}]$ and $P_{k|k}=\mathbb{E}[(x_k-x_{k|k})(x_k-x_{k|k})^T|\mathcal{I}_{1:k}]$, respectively.
\begin{figure}[thpb]
      \centering
      \includegraphics[scale=0.38]{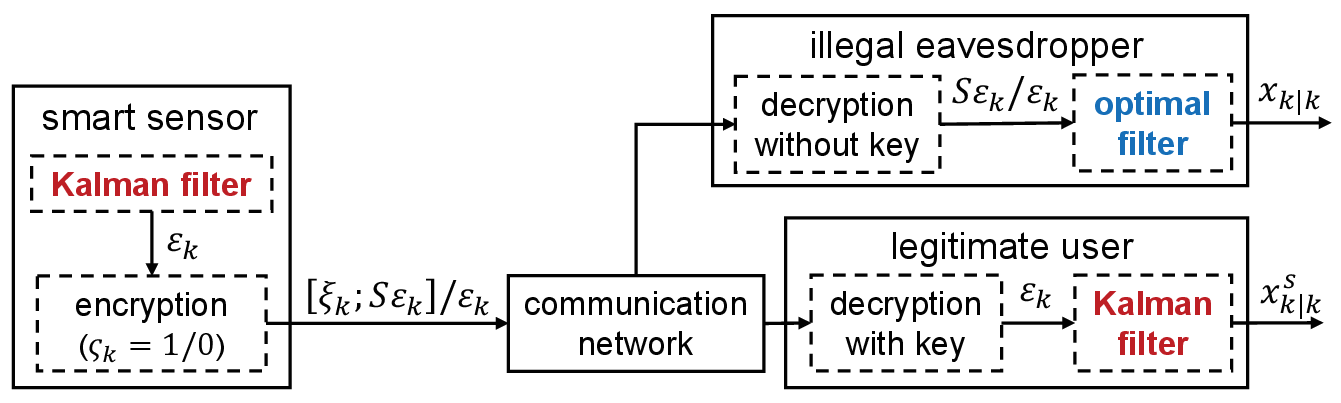}
      \caption{The framework of anti-eavesdropping state estimation.}
\end{figure}

In this paper, both stochastic and deterministic intermittent  strategies will be considered. Concretely, the stochastic intermittent strategy means that $\varsigma_k$ is determined by a given probability:
\begin{equation}\begin{aligned}
\mathrm{Pr}(\varsigma_k=1)=\varsigma,\ \mathrm{Pr}(\varsigma_k=0)=1-\varsigma.
\end{aligned}\end{equation}
The deterministic intermittent strategy means that $\varsigma_k$ is determined periodically:
\begin{equation}\begin{aligned}
\varsigma_k=f(k),
\end{aligned}\end{equation}
where $f(\cdot):\mathbb{N}_+\rightarrow\{0,1\}$ is a decision function with period $L$, i.e., $f(k)=f(k+L)$ for all $k\in\mathbb{N}_+$.

Then, the focus of this paper is summarized as follows.
\begin{itemize}
\item When the system (1) is stable, design encryption parameters $\varsigma$, $f(\cdot)$ and $\tilde{S}$ for both the stochastic and deterministic strategies to maximize the steady state of the MSE of the eavesdropper.
\item When the system (1) is unstable, design encryption parameters $\varsigma$, $f(\cdot)$ and $\tilde{S}$ for both the stochastic and deterministic strategies so that the MSE of the eavesdropper tends to infinity.
\end{itemize}

To facilitate the reader's understanding, Fig. 2 shows a high-level overview of the main results.
\begin{figure}[t]
      \centering
      \includegraphics[scale=0.5]{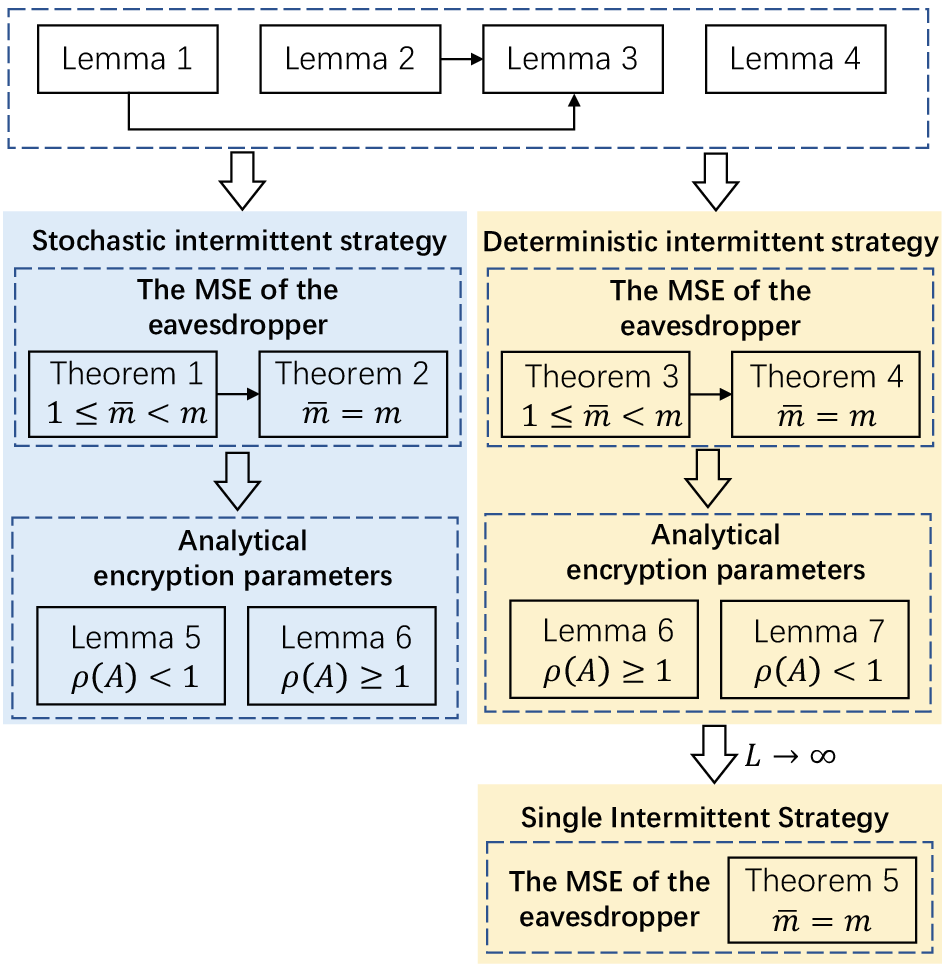}
      \caption{The relationships among main theorems and lemmas.}
\end{figure}

{\em Remark 1:} The proposed methods are different from the previous encryption-based methods in at least two aspects. 1) The scheduling policies in \cite{WANG2022110145} and \cite{HUANG2021109537} need to be derived by solving some optimization problems. Differently, all encryption parameters in this paper have analytical expressions. 2) For unstable systems, the work in \cite{Shang9882330} only verifies the unboundedness of the MSE of the eavesdropper by simulations without any rigorous theoretical support. By contrast, this paper proves the unboundedness based on the optimal filtering theory \cite{Optimalfiltering}.
%
\section{Main Results}
Before presenting the main results, we introduce the following lemmas.

\textbf{Lemma 1 \cite{Optimalfiltering}.} Consider some Gaussian random variables $x$, $s_1$, $s_2$, $\cdots$, $s_t$, where $\mathbb{E}[s_i]=0$, $\mathrm{Cov}(s_i,s_i)>0$ for $i=1,2,\cdots,t$ and $\mathrm{Cov}(s_i,s_j)=0$ if $i\neq j$. Then, one has $\mathbb{E}[x|s_{1:t}]=\mathbb{E}[x|s_{1:t-1}]+\mathrm{Cov}(x,s_t)\mathrm{Cov}^{-1}(s_t,s_t)s_t$ and $\mathrm{Cov}(x|s_{1:t})=\mathrm{Cov}(x|s_{1:t-1})-\mathrm{Cov}(x,s_t)\mathrm{Cov}^{-1}(s_t,s_t)\mathrm{Cov}^T(x,s_t)$, where $s_{1:i}\triangleq\{s_1,s_2,\cdots,s_i\}$.

\textbf{Lemma 2 \cite{Optimalfiltering}.} $\mathrm{Cov}(x_{k+i},\varepsilon_k)=A^iP^s_{k|k-1}C^T$ for $i\geq0$ and $k\geq1$. $\mathrm{Cov}(\varepsilon_k)=CP^s_{k|k-1}C^T+R$ for $k\geq1$.
%

\textbf{Lemma 3.} The MMSE estimate of the eavesdropper is given by
\begin{equation}\begin{aligned}
x_{k|k-1}=Ax_{k-1|k-1},
\end{aligned}\end{equation}
\begin{equation}\begin{aligned}
P_{k|k-1}=AP_{k-1|k-1}A^T+BQB^T,
\end{aligned}\end{equation}
\begin{equation}\begin{aligned}
x_{k|k}=&x_{k|k-1}+\varsigma_kP^s_{k|k-1}C^T_S(C_SP^s_{k|k-1}C^T_S+R_S)^{-1}S\varepsilon_k\\
&+(1-\varsigma_k)P^s_{k|k-1}C^T(CP^s_{k|k-1}C^T+R)^{-1}\varepsilon_k,
\end{aligned}\end{equation}
\begin{equation}\begin{aligned}
&P_{k|k}=P_{k|k-1}-\varsigma_k\Delta(S,k)-(1-\varsigma_k)\Delta(I,k),
\end{aligned}\end{equation}
where $C_S\triangleq SC$, $R_S\triangleq SRS^T$ if $S\neq0$, $R_S\triangleq I$ if $S=0$ and $\Delta(S,k)\triangleq P^s_{k|k-1}C^T_S(C_SP^s_{k|k-1}C^T_S+R_S)^{-1}C_SP^s_{k|k-1}$.

\textbf{Proof:} See Appendix A. $\square$

\textbf{Lemma 4.} For any $X\geq0$ and full-row-rank $L$ with appropriate dimensions, one has $XC^T(CXC^T+R)^{-1}CX\geq XC^TL^T(L(CXC^T+R)L^T)^{-1}LCX$.

\textbf{Proof:} See Appendix B. $\square$

Define the steady states
\begin{equation}\begin{aligned}
P^{s}_+\triangleq\lim_{k\rightarrow\infty}P^s_{k|k},\ P^s_{-}\triangleq\lim_{k\rightarrow\infty}P^s_{k|k-1},\ K^s\triangleq\lim_{k\rightarrow\infty}K^s_{k}.
\end{aligned}\end{equation}
Then the following assumption is necessary.

\textbf{Assumption 1.} There exists an $N\in\mathbb{N}_+\cup\{0\}$ such that the Kalman filter (2)-(6) reaches a steady state at $k=N$.

{\em Remark 2.} Since the Kalman filter defined by (2)-(6) converges to the steady state exponentially fast if the system (1) is stabilizable and detectable, a common and standard assumption in the Kalman filtering problem is $P_{0|0}=P^s_+$ \cite{WANG2022110145,Xu10054451,Tsiamis8758381,Leong8550317,Leong8543618,HUANG2021109537}. Evidently, $P_{0|0}=P^s_+$ is a special case of Assumption 1 with $N=0$. Moreover, although the initial condition is strict from a mathematical point of view, we can run the Kalman filter beforehand in practice until it reaches the steady state. Thus, Assumption 1 is acceptable in practical engineering. Meanwhile, Assumption 1 is of great help in deriving analytical solutions.
\subsection{Stochastic Intermittent Strategy}
When the stochastic intermittent strategy (10) is adopted, $P_{k|k}$ is a random variable because $\varsigma_k$ is random. Therefore, in this section we will discuss the property of $\mathbb{E}[P_{k|k}]$. It follows from Lemma 3 that the MMSE estimate $x_{k|k}$ of the eavesdropper is consistent with the standard Kalman filter (2)-(6) if $\varsigma=0$. Thus, we only discuss the case of $\varsigma>0$.

\textbf{Theorem 1:} Consider the system (1), the linear encryption scheme (7) with $1\leq\bar{m}<m$ and the stochastic intermittent strategy (10) with $\varsigma>0$. Under Assumption 1, one has the following results.
\begin{itemize}
\item[1.] When $\rho(A)<1$, the sequence $\{\mathbb{E}[P_{k|k}]\}$ converges to
\begin{equation}\begin{aligned}
\lim_{k\rightarrow\infty}\mathbb{E}[P_{k|k}]=&(1-\varsigma) P^s_++\varsigma\bar{P}_S,
\end{aligned}\end{equation}
where $\bar{P}_S$ is explicitly expressed by (22) and is the unique positive semi-definite solution to
\begin{equation}\begin{aligned}
X=AXA^T+BQB^T-\Delta(S),
\end{aligned}\end{equation}
where $\Delta(S)\triangleq P^s_-C^T_S(C_SP^s_-C^T_S+R_S)^{-1}C_SP^s_-$.
\item[2.] When $\rho(A)\geq1$, the sequence $\{\mathbb{E}[P_{k|k}]\}$ is unbounded if (18) does not have any positive semi-definite solution.
\end{itemize}

\textbf{Proof:} First, it will be proved by an induction that
\begin{equation}\begin{aligned}
\mathbb{E}[P_{k|k}]=(1-\varsigma)P^s_{k|k}+\varsigma\mathfrak{P}_{k|k},
\end{aligned}\end{equation}
where $\mathfrak{P}_{0|0}\triangleq P_{0|0}$ and
\begin{equation}\begin{aligned}
\mathfrak{P}_{k|k}=&A\mathfrak{P}_{k-1|k-1}A^T+BQB^T-\Delta(S,k).
\end{aligned}\end{equation}
It is trivial that $\mathbb{E}[\mathfrak{P}_{0|0}]=(1-\varsigma)P^s_{0|0}+\varsigma\mathfrak{P}_{0|0}$. Suppose that (19) holds at the $k-1$th moment. At the $k$th moment, since $\Delta(S,k)$ and $\Delta(I,k)$ are independent of $\varsigma_k$, one has
\begin{equation}\begin{aligned}
\mathbb{E}[P_{k|k}]=&A\mathbb{E}[P_{k-1|k-1}]A^T+BQB^T-\varsigma\Delta(S,k)\\
&-(1-\varsigma)\Delta(I,k)\\
=&(1-\varsigma)P^s_{k|k}+\varsigma\mathfrak{P}_{k|k}.
\end{aligned}\end{equation}
This completes the induction.
Since the system (1) is stabilizable and detectable, the sequence $\{P^s_{k|k}\}$ always converges. In this case, the convergence and boundedness of $\{\mathbb{E}[P_{k|k}]\}$ is determined by the sequence $\{\mathfrak{P}_{k|k}\}$.

1) Consider the case $\rho(A)<1$. Utilizing the convergence of the Kalman filter yields $\lim_{k\rightarrow\infty}\Delta(S,k)=\Delta(S)$. Therefore, it follows from (20) and the property of the discrete Lyapunov equation that
\begin{equation}\begin{aligned}
\lim_{k\rightarrow\infty}\mathfrak{P}_{k|k}=\sum^{\infty}_{i=0}A^i(BQB^T-\Delta(S))(A^T)^i,
\end{aligned}\end{equation}
which is the unique solution to (18). Moreover, according to Lemma 1 and the fact $\mathrm{Cov}(\varepsilon_i,\varepsilon_j)=0$ for all $i\neq j$, one can deduce that $\mathfrak{P}_{k|k}=\mathrm{Cov}(x_k|S\varepsilon_1,\cdots,S\varepsilon_k)\geq0$,
which means that the solution to (18) is positive semi-definite.

2) When $\rho(A)\geq1$, it follows from Assumption 1 that
\begin{equation}\begin{aligned}
\mathfrak{P}_{k+1|k+1}=&A\mathfrak{P}_{k|k}A^T+BQB^T-\Delta(S)
\end{aligned}\end{equation}
for $k\geq N$. Notice that $\mathcal{L}(S\varepsilon_1,\cdots,S\varepsilon_k)\subset \mathcal{L}(\varepsilon_1,\cdots,\varepsilon_k)=\mathcal{L}(z_1,\cdots,z_k)$. In this case, utilizing the optimality condition of the Kalman filter in the family of linear estimators, one knows $\mathfrak{P}_{k|k}\geq P^s_+,\ \forall k\geq N$. Then, consider the difference equation
\begin{equation}\begin{aligned}
X_{k+1}=AX_kA^T+BQB^T-\Delta(S)
\end{aligned}\end{equation}
with the initial point $X_0=P^s_+$. Utilizing Lemma 4 yields
\begin{equation}\begin{aligned}
X_1\geq& AP^s_+A^T+BQB^T-\Delta(I)= P^s_+=X_0.
\end{aligned}\end{equation}
With (25), it can be verified by an induction that $X_{k}\geq X_{k-1}$ for all $k\in\mathbb{N}_+$. In this case, one knows that (18) does not have any positive semi-definite solution is a sufficient condition for the unbounded sequence $\{X_k\}$. Moreover, it can be derived that $\mathfrak{P}_{k|k}\geq X_{k-N}$ for all $k\geq N$ by an induction. Therefore, $\{\mathfrak{P}_{k|k}\}$ will also be unbounded if (18) does not any positive semi-definite solution. The proof is completed. $\square$

%

Theorem 1 discusses the case of $1\leq\bar{m}<m$. The following theorem further discusses the case of $\bar{m}=m$.

\textbf{Theorem 2:} Consider the system (1), the encryption scheme (7) with $\bar{m}=m$ and the stochastic intermittent strategy (10) with $\varsigma>0$. Then, the following results hold.
\begin{itemize}
\item[1.] The sequence $\{\mathbb{E}[P_{k|k}]\}$ converges if and only if $\rho(A)<1$, and the limit point is given by (17).
\item[2.] The sequence $\{\mathbb{E}[P_{k|k}]\}$ is unbounded if and only if $\rho(A)\geq1$.
\end{itemize}

\textbf{Proof:} An argument similar to the one used in Theorem 1 shows that $\mathbb{E}[P_{k|k}]=(1-\varsigma)P^s_{k|k}+\varsigma\mathrm{Cov}(x_k)$, which implies that the boundedness and convergence of $\{\mathbb{E}[P_{k|k}]\}$ is determined by $\{\mathrm{Cov}(x_k)\}$. Moreover, it is trivial that $\mathrm{Cov}(x_k)=A\mathrm{Cov}(x_{k-1})A^T+BQB^T$, which is a discrete-time Lyapunov equation. When $\rho(A)<1$, it is clear that $\mathrm{Cov}(x_k)$ converges. When $\rho(A)\geq1$, $\lim_{k\rightarrow\infty}\mathrm{Cov}(x_k)\rightarrow\infty$ since $(A,B)$ is stabilizable. This completes the proof. $\square$

Theorems 1 and 2 state that the MSE of the eavesdropper converges for any encryption parameter if $\rho(A)<1$. Thus, we aim to design $S$ such that the steady state of the MSE of the eavesdropper is maximized given encryption consumption. Recall that $\varsigma$ and $\bar{m}$ denote the encryption probability and the dimension of the encrypted data, respectively, thus we can use the inequality $0\leq \varsigma\bar{m}\leq \mu_1$ to constrain the overall encryption consumption, where $0< \mu_1\leq m$. Moreover, the sensor cannot encrypt data with high dimension in one sampling period because of the limited computational capacity. Therefore, we also need to constrain the instantaneous encryption consumption with $\bar{m}\in\{1,2,\cdots,\mu_2\}$, where $1\leq \mu_2\leq m$. Based on the discussion above, the optimal encryption parameters are given by the optimization problem:
\begin{equation}\begin{aligned}
&\max_{\varsigma,\bar{m},S}\mathrm{Tr}((1-\varsigma)P^s_++\varsigma\bar{P}_S)\\
&\mathrm{s.t.}\ \mathrm{rank}(S)=m-\bar{m},\ 0\leq\varsigma\bar{m}\leq\mu_1,\\ &\ \ \ \ \ \bar{m}\in\{1,2,\cdots,\mu_2\}.
\end{aligned}\end{equation}
Then following lemma will solve the problem (26).

\textbf{Lemma 5:} An optimal solution to (26) belongs to the finite set $\{\varsigma(a),\bar{m}(a),S(a)|a\in\{1,2,\cdots,\mu_2\}\}$, where
\begin{equation}\begin{aligned}
\varsigma(a)=\mu_1/a,\ \bar{m}(a)=a,\ S(a)=[\varphi_1,\cdots,\varphi_{m-a}]^T,
\end{aligned}\end{equation}
$\varphi_i$ is the eigenvector of the matrix $(CP^s_-C^T+R)^{-1}CP^s_-WP^s_-C^T$ associated with the $i$th smallest eigenvalue and $W\triangleq\sum^{\infty}_{i=0}(A^{i})^TA^{i}$.

\textbf{Proof:} See Appendix C. $\square$

Furthermore, for $\rho(A)\geq1$, it follows from Theorem 1 that $\{\mathbb{E}[P_{k|k}]\}$ will be unbounded when (18) does not have any positive semi-definite solution. According to Theorem 4 in \cite{Shang9882330}, there exists an encryption matrix with $\bar{m}=1$ such that the requirement can be satisfied, i.e.,
\begin{equation}\begin{aligned}
\tilde{S}\in\left\{[x;X]\bigg|
\begin{array}{ll}
x\notin\mathrm{Ker}(w^HK^s),[x;X][x;X]^T=I,\\
x\in\mathbb{R}^{1\times m},X\in\mathbb{R}^{(m-1)\times m}
\end{array}
\right\},
\end{aligned}\end{equation}
where $w$ can be any vector satisfying $A^Tw=\lambda w$ with $|\lambda|\geq1$. The following lemma provides an analytical expression for $\tilde{S}$.

\textbf{Lemma 6:} If $\mathrm{Re}(w^HK^s)\neq0$, then an encryption matrix $\tilde{S}$ satisfying (28) is given by $\bar{S}\in\mathrm{Orth}(\mathrm{Re}(w^HK^s))$ and $S^T\in\mathrm{Orth}(\mathrm{Ker}(\mathrm{Re}(w^HK^s)))$, where $w$ can be any eigenvector of unstable eigenvalues of $A^T$. If $\mathrm{Re}(w^HK^s)=0$, then an encryption matrix $\tilde{S}$ satisfying (28) is given by $\bar{S}\in\mathrm{Orth}(\mathrm{Im}(w^HK^s))$ and $S^T\in\mathrm{Orth}(\mathrm{Ker}(\mathrm{Im}(w^HK^s)))$.

\textbf{Proof:} See Appendix D. $\square$

Then, we discuss the effect of $\varsigma$ and $S$ on the estimation performance of the eavesdropper. To this end, define two processes $\{\varsigma_{1,k}\}$ and $\{\varsigma_{2,k}\}$, where both $\varsigma_{1,k}$ and $\varsigma_{2,k}$ are binary variables taking 0 or 1. Moreover, define two matrices $S_1$ and $S_2$, which can be chosen as any full-row-rank matrix or $0$. When choosing $\varsigma_k=\varsigma_{i,k}$ and $S=S_i$ ($i=1,2$) as the encryption parameters, it follows from Lemma 3 that the error covariance of the eavesdropper (denoted as $P_{i,k|k}$) is given by
\begin{equation}\begin{aligned}
P_{i,k|k}=&AP_{i,k-1|k-1}A^T+BQB^T\\
&-(1-\varsigma_{i,k})\Delta(I,k)-\varsigma_{i,k}\Delta(S_i,k)
\end{aligned}\end{equation}
with initial point $P_{i,0|0}=P_{0|0}$.

\textbf{Corollary 1:} If $\varsigma_{i,k}$ is generated from the stochastic intermittent strategy (10) with $\mathrm{Pr}(\varsigma_{i,k}=1)=\tau_i$ for all $k\geq1$ and $i\in\{1,2\}$, the evolution (29) has the following properties.
\begin{itemize}
\item[1.] Keeping $S_1=S_2$, one has $\mathbb{E}[P_{1,k|k}]\leq \mathbb{E}[P_{2,k|k}]$ if $\tau_1\leq\tau_2$.
\item[2.] Keeping $\tau_1=\tau_2$, one has $\mathbb{E}[P_{1,k|k}]\leq \mathbb{E}[P_{2,k|k}]$ if $S_2^T\in \mathrm{R}(S_1^T)$.
\end{itemize}

\textbf{Proof:} See Appendix E. $\square$

{\em Remark 3.} Corollary 1 reveals the intuition that more encryption implies a higher privacy level in the stochastic intermittent strategy. Moreover, in Corollary 2, we will show that the deterministic intermittent strategy also obeys such an intuition.
\subsection{Deterministic Intermittent Strategy}
The deterministic intermittent strategy (11) will be considered in this section. Notice that $P_{k|k}$ is no longer a random variable in the deterministic intermittent strategy, so we will discuss $\{P_{k|k}\}$. Moreover, the MMSE estimate $x_{k|k}$ of the eavesdropper will be consistent with the Kalman filter (2)-(6) if $|f|=0$, where $|f|\triangleq\sum^L_{i=1}f(i)$. Thus, we only discuss the case of $|f|>0$.

\textbf{Theorem 3:} Consider the system (1), the linear encryption scheme (7) with $1\leq\bar{m}<m$ and the deterministic intermittent strategy (11) with $|f|>0$. Under Assumption 1, the following statements hold.
\begin{itemize}
\item[1.] When $\rho(A)<1$, the sequence $\{P_{k|k}\}$ converges periodically for all $S$, that is,
\begin{equation}\begin{aligned}
\lim_{i\rightarrow\infty}P_{iL+j|iL+j}=P(S,f,j)+&P^s_+,\\
&\forall j=1,\cdots,L,
\end{aligned}\end{equation}
where $P(S,f,j)$ is explicitly expressed by (38) and is the unique positive semi-definite solution to
\begin{equation}\begin{aligned}
X=A^LX&(A^L)^T+\sum^{L}_{i=1}f(i+j)\\
&\times A^{L-i}(\Delta(I)-\Delta(S))(A^{L-i})^T.
\end{aligned}\end{equation}
\item[2.] When $\rho(A)\geq1$, the sequence $\{P_{k|k}\}$ is unbounded if (19) does not have any positive semi-definite solution.
\end{itemize}

\textbf{Proof:} 1) First, it will be derived by an induction that
\begin{equation}\begin{aligned}
P_{k|k}=&P^s_{k|k}+\mathcal{D}(I,f,k,1)-\mathcal{D}(S,f,k,1),
\end{aligned}\end{equation}
where
\begin{equation}\begin{aligned}
\mathcal{D}(S,f,k,j)\triangleq\sum^k_{i=j}&f(i)\mathrm{Cov}(x_k,S\varepsilon_i)\mathrm{Cov}^{-1}(S\varepsilon_i)\\
&\times\mathrm{Cov}^T(x_k,S\varepsilon_i).
\end{aligned}\end{equation}
At the $1$st moment, (32) holds is clear. Suppose that (32) holds at the $k-1$th moment. Then, at the $k$th moment, one can derive from Lemma 2 and Lemma 3 that
\begin{align}
P_{k|k}=&A(P^s_{k-1|k-1}+\mathcal{D}(I,f,k-1,1)-\mathcal{D}(S,f,k-1,1))A^T\nonumber\\
&+BQB^T-\Delta(I,k)+f(k)(\Delta(I,k)-\Delta(S,k))\nonumber\\
=&P^s_{k|k}+\mathcal{D}(I,f,k,1)-\mathcal{D}(S,f,k,1).
\end{align}
This completes the induction.
According to the convergence of the Kalman filter, one knows $\lim_{\alpha\rightarrow\infty}P^s_{\alpha L+\beta|\alpha L+\beta}=P^s_+$ for all $\beta\in \mathbb{N}_+$. Based on the definition of $\mathcal{D}(S,f,k,j)$, one has
\begin{equation}\begin{aligned}
&\mathcal{D}(I,f,\alpha L+\beta,1)-\mathcal{D}(S,f,\alpha L+\beta,1)\\
=&A^L(\mathcal{D}(I,f,(\alpha-1)L+\beta,1)\\
&-\mathcal{D}(S,f,(\alpha-1)L+\beta,1))(A^L)^T\\
&+\mathcal{D}(I,f,\alpha L+\beta,(\alpha-1)L+\beta+1)\\
&-\mathcal{D}(S,f,\alpha L+\beta,(\alpha-1)L+\beta+1).
\end{aligned}\end{equation}
Then, it follows from the periodicity of $f(\cdot)$ and Lemma 2 that
\begin{equation}\begin{aligned}
&\mathcal{D}(S,f,\alpha L+\beta,(\alpha-1)L+\beta+1)\\
=&\sum^{L}_{i=1}f(i+\beta)A^{L-i}\Delta(S,(\alpha-1)L+\beta+i)(A^{L-i})^T.
\end{aligned}\end{equation}
In this case, utilizing the convergence of the Kalman filter yields
\begin{equation}\begin{aligned}
\lim_{\alpha\rightarrow\infty}\mathcal{D}(S,f,&\alpha L+\beta,(\alpha-1) L+\beta+1)\\
&=\sum^{L}_{i=1}f(i+\beta)A^{L-i}\Delta(S)(A^{L-i})^T.
\end{aligned}\end{equation}
Since $\rho(A^L)=\rho^L(A)<1$, according to the property of the discrete Lyapunov equation, one has
\begin{equation}\begin{aligned}
&\lim_{\alpha\rightarrow\infty}\mathcal{D}(I,f,\alpha L+\beta,1)-\mathcal{D}(S,f,\alpha L+\beta,1)\\
=&\sum^\infty_{j=0}A^{jL}\left(\sum^{L}_{i=1}f(i+\beta)A^{L-i}(\Delta(I)-\Delta(S))(A^{L-i})^T\right)\\
&\ \ \ \ \times(A^{jL})^T,
\end{aligned}\end{equation}
which is the unique solution to (31). Moreover, it follows from Lemma 4 that (38) is positive semi-definite.

2) Then, we consider the case where $\rho(A)\geq1$. According to (32), one can deduce that
for any decision function $f(\cdot)$, the covariance $P_{\alpha L+\beta|\alpha L+\beta}$ satisfies
\begin{equation}\begin{aligned}
P_{\alpha L+\beta|\alpha L+\beta}\geq& \bar{P}_{\alpha L+\beta|\alpha L+\beta}\\
=&P^s_{\alpha L+\beta|\alpha L+\beta}+\mathcal{D}(I,\bar{f},\alpha L+\beta,1)\\
&-\mathcal{D}(S,\bar{f},\alpha L+\beta,1),
\end{aligned}\end{equation}
where $\bar{f}(\cdot):\mathbb{N}_+\rightarrow\{0,1\}$ satisfies
\begin{equation}\begin{aligned}
\bar{f}(k)
=\left\{ \begin{array}{l}
1,\ \forall k=iL+\gamma,\ i\geq0\\
0,\ \mathrm{otherwise}
\end{array} \right.
\end{aligned}\end{equation}
with $1\leq \gamma\leq L$ and $f(\gamma)=1$. It remains to show that $\{\bar{P}_{k|k}\}$ is unbounded. Under Assumption 1, there exists a positive integer $\alpha$ such that the Kalman filter converges for all $k\geq N=\alpha L+\gamma$. In this case, invoking the definition of $\mathcal{D}(S,f,i,j)$ yields
\begin{equation}\begin{aligned}
&\mathcal{D}(I,\bar{f},\alpha_1L+\gamma_1,1)-\mathcal{D}(S,\bar{f},\alpha_1L+\gamma_1,1)\\
=&A^{(\alpha_1-\alpha)L+\gamma_1-\gamma+1}(\mathcal{D}(I,\bar{f},\alpha L+\gamma-1,1)\\
&-\mathcal{D}(S,\bar{f},\alpha L+\gamma-1,1))(A^{(\alpha_1-\alpha)L+\gamma_1-\gamma+1})^T\\
&+A^{L+\gamma_1-\gamma}\sum^{\alpha_1-\alpha-1}_{i=0}A^{iL}(\Delta(I)-\Delta(S))(A^T)^{iL}\\
&\times(A^{L+\gamma_1-\gamma})^T,
\end{aligned}\end{equation}
where $\alpha_1\geq \alpha+1$ and $\gamma_1<\gamma$.
Next, we will prove that (41) is unbounded when $\alpha_1\rightarrow\infty$. When $\gamma_1<\gamma$, one can deduce that
\begin{equation}\begin{aligned}
&\sum^{L-1}_{j=0}A^jA^{L+\gamma_1-\gamma}\sum^{\alpha_1-\alpha-1}_{i=0}A^{iL}(\Delta(I)-\Delta(S))(A^{iL})^T\\ &\times(A^{L+\gamma_1-\gamma})^T(A^j)^T\\
=&\sum^{(\alpha_1-\alpha+1)L-1+\gamma_1-\gamma}_{i=0}A^{i}(\Delta(I)-\Delta(S))(A^i)^T\\
&-\sum^{L+\gamma_1-\gamma-1}_{i=0}A^{i}(\Delta(I)-\Delta(S))(A^i)^T.
\end{aligned}\end{equation}
Consider the following difference equations
\begin{equation}\begin{aligned}
X_{k+1}=\mathfrak{G}\circ\mathfrak{L}(X_k)
\end{aligned}\end{equation}
\begin{equation}\begin{aligned}
Y_{k+1}=\mathfrak{L}(Y_k)-\Delta(S)
\end{aligned}\end{equation}
with initial point $X_0=Y_0=P^s_+$, where $\mathfrak{L}(X)\triangleq AXA^T+BQB^T$ and $\mathfrak{G}(X)\triangleq X-XC^T(CXC^T+R)CX$. According to the convergence of the Kalman filter, one knows that (43) is a fixed point equation. In this case, one obtains
\begin{equation}\begin{aligned}
&X_{k+1}=\mathfrak{G}\circ\mathfrak{L}(X_k)\\
=&A^{k+1}P^s_+(A^{k+1})^T+\sum^{k}_{i=0}A^iBQB^T(A^i)^T\\
&-\sum^{k}_{i=0}A^{i}\Delta(I)(A^{i})^T.
\end{aligned}\end{equation}
Moreover, by recursively applying (44), one has
\begin{equation}\begin{aligned}
Y_{k+1}=&A^{k+1}P^s_+(A^{k+1})^T+\sum^{k}_{i=0}A^iBQB^T(A^i)^T\\
&-\sum^{k}_{i=0}A^{i}\Delta(S)(A^{i})^T.
\end{aligned}\end{equation}
Combining (45) and (46) yields
\begin{equation}\begin{aligned}
Y_{k+1}-X_{k+1}=\sum^{k}_{i=0}A^{i}(\Delta(I)-\Delta(S))(A^{i})^T.
\end{aligned}\end{equation}
Since $\{X_k\}$ is a bounded sequence, $\{Y_{k+1}-X_{k+1}\}$ is bounded if and only if $\{Y_{k+1}\}$ is bounded. With this fact, one knows from the proof of Theorem 1 that $\{Y_{k+1}-X_{k+1}\}$ will be unbounded if (18) does not have any positive semi-definite solution. Note that the second term at the right-hand side of (42) is a constant matrix, thus one has
\begin{equation}\begin{aligned}
\lim_{\alpha_1\rightarrow\infty}\mathrm{Tr}(&\sum^{L-1}_{j=0}A^jA^{L+\gamma_1-\gamma}\sum^{\alpha_1-\alpha-1}_{i=0}A^{iL}(\Delta(I)-\Delta(S))\\
&\times (A^T)^{iL}(A^{L+\gamma_1-\gamma})^T(A^j)^T)\rightarrow\infty.
\end{aligned}\end{equation}
Notice that $\sum^{L-1}_{j=0}(A^T)^jA^j>0$, which indicates that there exists a positive number $p>0$ such that
\begin{equation}\begin{aligned}
\lim_{\alpha_1\rightarrow\infty}p \mathrm{Tr}(&A^{L+\gamma_1-\gamma}\sum^{\alpha_1-\alpha-1}_{i=0}A^{iL}(\Delta(I)-\Delta(S))(A^T)^{iL}\\ &\times(A^{L+\gamma_1-\gamma})^T)\rightarrow \infty.
\end{aligned}\end{equation}
Meanwhile, it follows from Lemma 4 that
\begin{equation}\begin{aligned}
&\Delta(I)-\Delta(S)\geq0,\\
&\mathcal{D}(I,\bar{f},\alpha L+\gamma-1,1)-\mathcal{D}(S,\bar{f},\alpha L+\gamma-1,1)\geq0.
\end{aligned}\end{equation}
Thus, the sequence $\{\bar{P}_{\alpha_1L+\gamma_1|\alpha_1L+\gamma_1}\}$ is unbounded when $\gamma_1<\gamma$. Moreover, if $\gamma_1\geq \gamma$, $\{\bar{P}_{\alpha_1L+\gamma_1|\alpha_1L+\gamma_1}\}$ can be proved to be unbounded based on the same analysis procedure, which is omitted for brevity. The proof is completed. $\square$

Theorem 3 discusses the case where $1\leq\bar{m}<m$. Next, Theorem 4 will discuss the case where $\bar{m}=m$.

\textbf{Theorem 4:} Consider the system (1), the linear encryption scheme (7) with $S=0$ and the deterministic intermittent strategy (11) with $|f|>0$. Under Assumption 1, the following statements hold.
\begin{itemize}
\item[1)] The sequence $\{P_{k|k}\}$ converges periodically if and only if $\rho(A)<1$. Meanwhile, the limit point is given by (31).
\item[2)] The sequence $\{P_{k|k}\}$ is unbounded if and only if $\rho(A)\geq1$.
\end{itemize}

\textbf{Proof:} Similar to the proof of Theorem 3, one has
\begin{equation}\begin{aligned}
&P_{\alpha L+\beta|\alpha L+\beta}\\
=&P^s_{\alpha L+\beta|\alpha L+\beta}+\mathcal{D}(I,f,\alpha L+\beta,1)\\
=&P^s_{\alpha L+\beta|\alpha L+\beta}+A^L\mathcal{D}(I,f,(\alpha-1)L+\beta,1)(A^L)^T\\
&+\mathcal{D}(I,f,\alpha L+\beta,(\alpha-1)L+\beta+1).
\end{aligned}\end{equation}

1) When $\rho(A)<1$, the result is straightforward from Theorem 3.

2) When $\rho(A)\geq1$, it follows from the Popov–Belevitch–Hautus (PBH) test that
\begin{equation}\begin{aligned}
\lim_{k\rightarrow\infty}A^{k+1}P^s_+(A^{k+1})^T+\sum^{k}_{i=0}A^iBQB^T(A^i)^T\rightarrow\infty.
\end{aligned}\end{equation}
Since (43) is a fixed point equation, one has $\lim_{k\rightarrow\infty}X_k=P^s_+$. In this case, it can be deduced from (45) and (52) that $\lim_{k\rightarrow\infty}\sum^{k}_{i=0}A^{i}\Delta(I)(A^{i})^T\rightarrow\infty$.
Then, under Assumption 1, the second statement can be proved along similar lines as in the proof of Theorem 3. The proof is completed. $\square$

Next, we discuss how to maximize the steady state of the MSE of the eavesdropper with a given encryption consumption when $\rho(A)<1$. Note that $|f|$ denotes the number of encryptions performed in a period and $L$ denotes the length of a period. Thus, $|f|/L$ denotes the encryption frequency. Moreover, one can know from (30) that $\{P_{iL+j|iL+j}\}$ converges to different steady state for different $j=1,\cdots,L$, thus their average value will be used as the objective value. With the discussion above, similar to (26), the optimal encryption parameters are given by the optimization problem
\begin{equation}\begin{aligned}
&\max_{\frac{|f|}{L},\bar{m},S} \sum^{L}_{i}\frac{1}{L}\mathrm{Tr}(P^s_++P(S,f,i))\\
&\mathrm{s.t.}\ \mathrm{rank}(S)=m-\bar{m},\ 0\leq\frac{|f|}{L}\bar{m}\leq\mu_3,\\ &\ \ \ \ \ \bar{m}\in\{1,2,\cdots,\mu_4\}.
\end{aligned}\end{equation}
The following lemma will solve the problem (53).

\textbf{Lemma 7:} An optimal solution to (53) belongs to the finite set $\{\frac{|f_a|}{L_a},\bar{m}(a),S(a)|a\in\{1,2,\cdots,\mu_4\}\}$, where
\begin{equation}\begin{aligned}
\frac{|f_a|}{L_a}=\frac{\mu_3}{a},\ \bar{m}(a)=a,\ S(a)=[\varphi_1,\cdots,\varphi_{m-a}]^T.
\end{aligned}\end{equation}

\textbf{Proof:} See Appendix F. $\square$

The following corollary illustrates the effect of $f(\cdot)$ and $S$ on the estimation performance of the eavesdropper. 

\textbf{Corollary 2:} When $\varsigma_{i,k}$ is generated from the deterministic intermittent strategy (11) with decision function $f_i(\cdot)$ (with period $L_i$) for $i\in\{1,2\}$, then the evolution (29) has the following properties.
\begin{itemize}
\item[1.] Keeping $S_1=S_2$, one has $P_{1,k|k}\leq P_{2,k|k}$ if $f_1(k)\leq f_2(k)$ for $k\in\mathbb{N}_+$.
\item[2.] Keeping $S_1=S_2$, then $\lim_{i\rightarrow\infty}\frac{1}{L_1}\sum^{L_1}_{j=1}P_{1,iL+j|iL+j}\leq\lim_{i\rightarrow\infty}\frac{1}{L_2}\sum^{L_2}_{j=1}P_{2,iL+j|iL+j}$ if $|f_1|/L_1\leq |f_2|/L_2$ and $\rho(A)<1$.
\item[3.] Keeping $f_1(k)=f_2(k)$ for $k\in\mathbb{N}_+$, one has $P_{1,k|k}\leq P_{2,k|k}$ if $S^T_2\in \mathrm{R}(S^T_1)$.
\end{itemize}

\textbf{Proof:} See Appendix G. $\square$

The following corollary shows the relationship between the stochastic and deterministic intermittent strategies.

\textbf{Corollary 3:} Consider the case where $\rho(A)<1$, then the optimization problems (26) and (53) have the same optimal objective values and decision variables if $\mu_1=\mu_3$ and $\mu_2=\mu_4$.

\textbf{Proof:} See Appendix H. $\square$

\subsection{Single Intermittent Strategy}
This section will discuss the single intermittent encryption strategy
\begin{equation}\begin{aligned}
\varsigma_k=
\left\{ \begin{array}{l}
1,\ k=\delta,\\
0,\ \mathrm{otherwise},
\end{array} \right.
\end{aligned}\end{equation}
where $\delta\in\mathbb{N}_+$. The single intermittent strategy (55) extends the case of the deterministic intermittent strategy (11) with period $L\rightarrow\infty$. However, since the analysis methodology for the deterministic strategy is not applicable if $L\rightarrow\infty$, the single intermittent strategy will be discussed separately as follows. As only one encryption is performed by (55), even encrypting the entire $\varepsilon_k$ does not cause much encryption consumption. Thus, we only discuss the case $\bar{m}=m$.

\textbf{Theorem 5:} Consider the system (1), the encryption scheme (7) with $\bar{m}=m$ and the single intermittent strategy (55). Then, the following statements hold.
\begin{itemize}
\item[1.] When $\rho(A)<1$, any single intermittent strategy cannot degrade the MSE of the eavesdropper, that is, $\lim_{k\rightarrow\infty}P_{k|k}=P^s_+$.
\item[2.] When $\rho(A)=1$, the sequence $\{P_{k|k}\}$ is unbounded only if the unstable Jordan blocks of $A$ are not all scalars.
\item[3.] When $\rho(A)>1$, the sequence $\{P_{k|k}\}$ is unbounded if there exists $Y>0$ and $Z$ such that
    \begin{equation}\begin{aligned}
\begin{bmatrix}
{Y} & {YA^T-ZCP^s_{\delta|\delta-1}} & {Y}\\
{(YA^T-ZCP^s_{\delta|\delta-1})^T} & {Y} & {0}\\
{Y} & {0} & {I}
\end{bmatrix}\geq0.
\end{aligned}\end{equation}
\end{itemize}

\textbf{Proof:} See Appendix I. $\square$

{\em Remark 4:} The intermittent strategies discussed in sections III-A and III-B require infinite encryption over an infinite time horizon. Differently, the single intermittent strategy (55) has negligible encryption consumption over an infinite time domain since it performs the encryption scheme (7) only once.
\section{Discussion on Practical Issues}
\subsection{Differences between the proposed stochastic and deterministic intermittent strategies}
When $\rho(A)\geq1$ and the encryption matrix $S$ is chosen according to Lemma 6, both the stochastic and deterministic intermittent strategies can leave the eavesdropper's MSE unbounded with any positive encryption frequency. When $\rho(A)<1$, Corollary 3 proves that the stochastic and deterministic intermittent strategies have the same effect on the eavesdropper's steady-state performance. Therefore, theoretically, there is no essential difference between the stochastic and deterministic intermittent strategies. However, there are some remarkable differences between the two strategies in practice.

We will analyze the possible advantage of the deterministic intermittent strategy in practice through an intuitive example. Suppose that the user expects the encryption frequency to be 0.1. Then, the encryption probability $\varsigma$ in the stochastic intermittent strategy (10) needs to be set to $\varsigma=0.1$. Accordingly, the decision function $f(\cdot)$ in the deterministic intermittent strategy (11) can be set to
    \begin{equation}
    \begin{aligned}
    f(k)
    =\left\{ \begin{array}{l}
    1,\ \forall k=10i+1,\ i=0,1,2,\cdots,\\
    0,\ \mathrm{otherwise}.
    \end{array} \right.
    \nonumber
    \end{aligned}
    \end{equation}
If the stochastic intermittent strategy is employed, the encryption may be required over several successive sampling periods. This would result in the computation being piled up in a time horizon. On the contrary, if the deterministic intermittent strategy is employed, each encryption is separated by 10 sampling periods, thus avoiding the problem of computation crowding.

Then, we will analyze the possible advantages of the stochastic intermittent strategy in practice. In practice, the eavesdropper does not always know whether the data is encrypted or not since the ciphertext $\xi_k$ and the plaintext $\bar{S}\varepsilon_k$ have the same dimension. Therefore, the eavesdropper needs to detect $\varsigma_k$ by some encryption detection methods. In general, the encryption detection can be accomplished by testing the statistical properties of the data (entropy and distribution) and the potential identification character. Note that the accuracy rate of the detection cannot be $100\%$. As a result, the eavesdropper may classify the encrypted message as an un-encrypted one, thus inputting problematic data into the estimator; or, classify the un-encrypted message as an encrypted one, thus losing useful information. These detection errors can lead to performance degradation of the eavesdropper. In the deterministic intermittent strategy, encryption is performed periodically, and this regularity may be exploited by the eavesdropper to obtain $\varsigma_k$ accurately. In contrast, the stochastic intermittent strategy is completely random and does not have any regularity. This randomness greatly increases the difficulty of the encryption detection and therefore contributes to the user's privacy.

The discussion above states that in practice, the deterministic intermittent strategy may have a better advantage in terms of computational efficiency, and the stochastic intermittent strategy may have a better advantage in terms of the privacy. Moreover, although the eavesdropper may not be able to know $\varsigma_k$ in practice, this paper still assumes that the eavesdropper has access to $\varsigma_k$. This makes the eavesdropper strong. From the user's perspective, the eavesdropper knowing $\varsigma_k$ is a conservative assumption, which has better robustness.
\subsection{Rounding errors}
This section analyzes the effect of rounding errors on the estimation performance of the user. Note that there may exist rounding errors when the user accesses $\varepsilon_k$ by inverting $\tilde{S}$. In this case, the available message can be represented as $(I+\Theta)\varepsilon_k$, where $\Theta$ comes from the rounding error in inverting $\tilde{S}$. We denote the estimate with the rounding error as $x^r_{k|k}$ to distinguish it from the ideal MMSE estimate $x^s_{k|k}$. When the rounding error exists (and is not detected by the user), the estimate $x^r_{k|k}$ is given by
    \begin{equation}
    \begin{aligned}
    x^r_{k|k-1}=Ax^r_{k-1|k-1},
    \end{aligned}
    \end{equation}
    \begin{equation}
    \begin{aligned}
    P^s_{k|k-1}=AP^s_{k-1|k-1}A^T+BQB^T,
    \end{aligned}
    \end{equation}
    \begin{equation}
    \begin{aligned}
    K^s_k=P^s_{k|k-1}C^T(CP^s_{k|k-1}C^T+R)^{-1},
    \end{aligned}
    \end{equation}
    \begin{equation}
    \begin{aligned}
    x^r_{k|k}=x^r_{k|k-1}+K^s_k(I+\Theta)\varepsilon_k,
    \end{aligned}
    \end{equation}
    \begin{equation}
    \begin{aligned}
    P^s_{k|k}=&(I-K^s_kC)P^s_{k|k-1},
    \end{aligned}
    \end{equation}
    where $x^r_{0|0}=x_{0|0}$. A straightforward induction shows that
    \begin{equation}
    \begin{aligned}
    x^r_{k|k}=x^s_{k|k}+e_k,
    \end{aligned}
    \end{equation}
where $e_k=Ae_{k-1}+K^s_{k}\Theta\varepsilon_{k}$ with $e_0=0$. It can be seen from (62) that the estimate $x^r_{k|k}$ deviates from the ideal MMSE estimate $x^s_{k|k}$ by $e_k$. The covariance of $e_k$ can be given by
    \begin{equation}
    \begin{aligned}
    \mathrm{Cov}(e_k)=A\mathrm{Cov}(e_{k-1})A^T+K^s_{k}\Theta\mathrm{Cov}(\varepsilon_{k})\Theta^T(K^s_{k})^T.
    \end{aligned}
    \end{equation}
    Meanwhile, note that $\lim_{k\rightarrow\infty}\mathrm{Cov}(\varepsilon_{k})=CP^s_-C^T+R$ and $\lim_{k\rightarrow\infty}K^s_k=K^s$. Thus, if $\rho(A)<1$,
    \begin{equation}
    \begin{aligned}
    &\lim_{k\rightarrow\infty}\mathrm{Cov}(e_k)\\
    =&\sum^{\infty}_{i=0}A^i(K^s\Theta(CP^s_-C^T+R)\Theta^T(K^s)^T)(A^T)^i.
    \end{aligned}
    \end{equation}
    This means that the rounding error makes the user's estimate error larger (but still bounded) in the stable system. On the other hand, one can know from (63) that, if $\rho(A)\geq1$, the covariance $\mathrm{Cov}(e_k)$ is likely to be amplified by $A$ to infinity, thus destroying the user's estimator.

Next, we discuss how to solve the issue induced by the rounding error. Since $\{\varepsilon_k\}$ is an orthogonal Gaussian stochastic process with known mean and covariance, we can determine whether the rounding error exists by testing the theoretical and actual statistical properties of the available message. Alternatively, we can directly test if $\tilde{S}\tilde{X}=I$, where $\tilde{X}$ is the inverse of $\tilde{S}$ computed by the user ($\tilde{X}$ may be ill-conditioned). If the actual statistical property is significantly different from the theoretical one, then the rounding error exists and has a significant effect on the user. Similarly, if $\tilde{S}\tilde{X}$ is significantly different from $I$, then the rounding error exists and has a significant effect on the user. In this case, we shall recalculate $\tilde{S}^{-1}$ by some software or computers with high accuracy. Even if the test does not reveal problems, we still recommend utilizing computers and software with high accuracy to compute $\tilde{S}^{-1}$, as this is an off-line operation with little overhead. After the test and recalculation, the rounding error can be almost eliminated. However, the rounding error cannot be eliminated completely due to the internal quantization within the computer (but they will be very small after the previous steps).

Finally, we analyze how to deal with the small rounding error (i.e., $\Theta$ is very small). In the stable system, it can be seen from (64) that $\mathrm{Cov}(e_k)$ is also small if $\Theta$ is small. In this case, there is no need for further processing. In the unstable system, even a small $\Theta$ may cause the estimate $x^r_{k|k}$ to deviate completely from the ideal MMSE estimate $x^s_{k|k}$. To solve this problem, we can have the sensor occasionally transmit fully encrypted $x^s_{k|k}$ to the user.
The user then decrypts the received message to obtain $x^s_{k|k}$ and resets the estimator with $x^s_{k|k}$, i.e., $x^r_{k|k}$ is replaced by $x^s_{k|k}$. This prevents the rounding error from being continually amplified by $A$. Note that, $x^r_{k|k}$ will not differ too much from $x^s_{k|k}$ even if $x^s_{k|k}$ is encrypted with a low frequency since $\Theta$ is very small. Therefore, this does not increase the encryption consumption too much. In addition, $\mathrm{Cov}(e_k)$ is not necessarily unbounded if $(A,K^s_k\Theta)$ is not uniformly stabilizable. For example, consider the difference equation
\begin{equation}
\begin{aligned}
X_k=AX_{k-1}A^T+Q,
\end{aligned}
\end{equation}
where $A=\mathrm{diag}(2,0.1)$ and $X_0=Q=\mathrm{diag}(0,1)$. Clearly, $\rho(A)=2\geq1$, but the limit is given by
\begin{equation}
\begin{aligned}
\lim_{k\rightarrow\infty}X_k=\mathrm{diag}(0,\sum^{\infty}_{i=0}0.01^i)=\mathrm{diag}(0,\frac{1}{0.99}).
\end{aligned}
\end{equation}
\section{Simulation}
Consider the damped mass-spring system \cite{Hu9705534}:
\begin{equation}\begin{aligned}
\frac{d x_t}{d t}=&
\begin{bmatrix}
{0} & {0} & {1} & {0}\\
{0} & {0} & {0} & {1}\\
{-\frac{k_1}{m_1}} & {\frac{k_1}{m_1}} & {-\frac{c_1}{m_1}} & {\frac{c_1}{m_1}}\\
{\frac{k_1}{m_2}} & {-\frac{k_1+k_2}{m_2}} & {\frac{c_1}{m_2}} & {-\frac{(c_1+c_2)}{m_2}}
\end{bmatrix}x_t\\
&+
\begin{bmatrix}
{0} & {0}\\
{0} & {0}\\
{\frac{1}{m_1}} & {0}\\
{0} & {\frac{1}{m_2}}
\end{bmatrix}
\begin{bmatrix}
{w_{1,t}}\\
{w_{2,t}}
\end{bmatrix},
\nonumber
\end{aligned}\end{equation}
where $x_t=[x_{1,t};\dot{x}_{1,t};x_{2,t};\dot{x}_{2,t}]$, $x_{i,t}$ and $\dot{x}_{i,t}$ and represent the displacement and velocity of the $i$th ($i=1,2$) object, respectively, $m_i$ represents the mass of the $i$th object, $k_i$ and $c_i$ are respectively the $i$th spring factor and damping factor. In this simulation, the parameters are chosen as $m_1=1$, $m_2=2$, $k_1=20$, $k_2=1$. We then discretize the continuous-time system above using the ``c2d" function in MATLAB. The sampling period is $\Delta t=0.1$s. Additionally, the measurement equation is given by
\begin{equation}
\begin{aligned}
z_{k\Delta t}=\begin{bmatrix}
{1} & {0} & {0} & {0}\\
{0} & {1} & {0} & {0}
\end{bmatrix}x_{k\Delta t}+v_{k\Delta t},\nonumber
\end{aligned}
\end{equation}
where $v_{k\Delta t}$ is a white Gaussian noise with covariance $0.25I$.

First, let us set the damping $c_1=c_2=-1$, in which case $\rho(A)=1.09$. Fig. 3 shows the MSEs of the eavesdropper and the user when different intermittent strategies are employed, where the encryption probability $\varsigma=0.1$, the decision function $f(k)=1$ only if $k=10i+1$ and the encryption moment $\delta=1$ (it can be verified by ``feasp" function in MATLAB that (56) is feasible when $\delta=1$). As we can see, each strategy can lead to an unbounded MSE of the eavesdropper.
Then set damping $c_1=c_2=1$, in which case $\rho(A)=0.98$. We consider a low encryption consumption case, i.e., $\mu_1=\mu_3=0.2$ and $\mu_2=\mu_4=2$. According to Lemma 5 and Lemma 7, one can obtain that the optimal encryption parameters are $\bar{m}=1$ and $\varsigma=\frac{|f|}{L}=0.2$. Fig. 4 shows the MSEs of the user and the eavesdropper, from which we see that intermittent strategies effectively reduce the estimation performance of the eavesdropper. Also, it is observed from Fig. 4 that the single strategy does not reduce the estimation performance of the eavesdropper, which is consistent with our conclusion in Theorem 5.
\begin{figure}[t]
      \centering
      \includegraphics[scale=0.65]{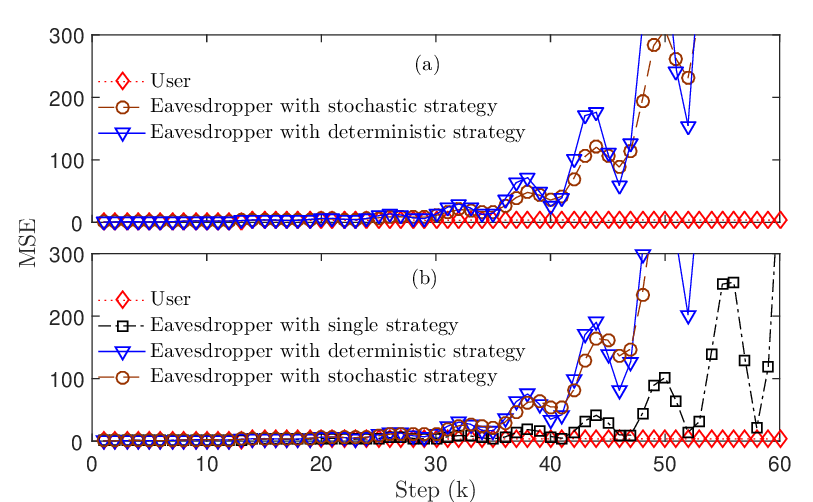}
      \caption{When the system is unstable, MSEs of the user and the eavesdropper under different intermittent strategies. In (a), $\bar{m}=1$ and $\tilde{S}$ is obtained by Lemma 6. In (b), $\bar{m}=2$, i.e., $\varepsilon_k$ is encrypted completely.}
\end{figure}
\begin{figure}[t]
      \centering
      \includegraphics[scale=0.65]{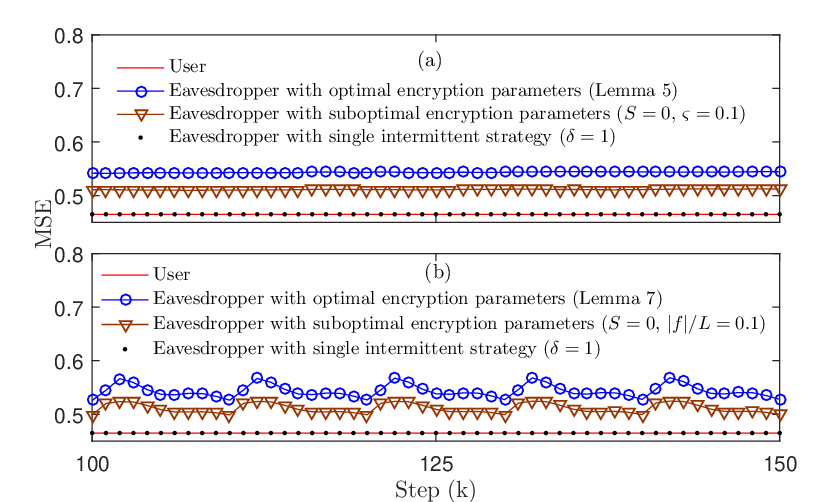}
      \caption{When the system is stable, MSEs of the user and the eavesdropper under different intermittent strategies. (a) Stochastic intermittent strategy. (b) Deterministic intermittent strategy.}
\end{figure}

To elaborate on the advantage of the proposed methods in terms of the encryption consumption, we compare them with the methods in \cite{WANG2022110145} and \cite{Shang9882330}. Here, the encryption consumption is portrayed by {\em encryption frequency}$\times${\em  dimension of the encrypted message}. Considering the case $c_1=c_2=-1$, the privacy levels of different methods are presented in Fig. 5. This figure shows that both the proposed methods and the method in \cite{Shang9882330} leave the eavesdropper's MSE unbounded. However, the encryption consumption of the method in \cite{Shang9882330} is $1\times 1=1$, which is much larger than the proposed methods. Additionally, the encryption frequency of the method in \cite{WANG2022110145} is set to 0.05 to keep the encryption consumption consistent. As can be seen from Fig. 5, the method in \cite{WANG2022110145} does not leave the eavesdropper's MSE unbounded. Thus, the proposed methods provide a better privacy level while maintaining the same encryption consumption as the method in \cite{WANG2022110145}. In particular, the proposed single intermittent strategy makes the eavesdropper's MSE tend to infinity with almost negligible encryption consumption (note that $1/\infty=0$), further illustrating the advantage of the proposed methods in terms of the encryption consumption.
\begin{figure}[t]
      \centering
      \includegraphics[scale=0.65]{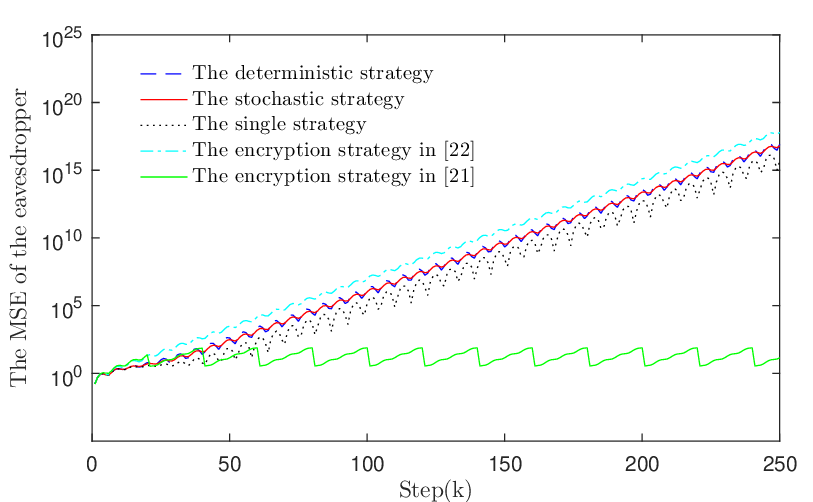}
      \caption{The privacy levels of the proposed encryption strategies, the encryption strategy in [21] and the encryption strategy in [22].}
\end{figure}
\section{Conclusion}
To reduce the encryption consumption in anti-eavesdropping estimation problems, stochastic, deterministic, and single intermittent strategies were proposed in this paper based on different problem settings. For unstable systems, the proposed stochastic and deterministic intermittent strategies can make the MSE of the eavesdropper unbounded with any encryption frequency, thus greatly reducing the consumption of the encryption algorithm. For stable systems, the proposed stochastic and deterministic intermittent strategies do not make the MSE of the eavesdropper unbounded. In this case, we gave an analytical expression of the encryption matrix that maximizes the steady state of the MSE of the eavesdropper. The single intermittent strategy, whose encryption consumption is almost negligible, is shown to be unusable in stable systems but can let the eavesdropper's MSE unbounded in unstable systems if a specified LMI is feasible.
\section*{Appendix}
\subsection{Proof of Lemma 3}
According to the orthogonality of the innovation sequence $\{\varepsilon_k\}$ (see Section 5.3 of \cite{Optimalfiltering}), one knows $\mathrm{Cov}(y_i,y_j)=0$ for $i\neq j$. Given any $\{\varsigma_1,\cdots,\varsigma_k\}\in\{0,1\}^k$. If $S=0$ and $\varsigma_k=1$, one has $y_k=0$, i.e., no information is provided. In this case, it is trivial that
\begin{equation}\begin{aligned}
x_{k|k}=x_{k|k-1},
\end{aligned}\end{equation}
\begin{equation}\begin{aligned}
P_{k|k}=P_{k|k-1}.
\end{aligned}\end{equation}
If $S=0$ and $\varsigma_k=0$, utilizing Lemma 2 yields
\begin{equation}\begin{aligned}
x_{k|k}=&x_{k|k-1}+\mathrm{Cov}(x_k,\varepsilon_k)\mathrm{Cov}^{-1}(\varepsilon_k)\varepsilon_k,
\end{aligned}\end{equation}
\begin{equation}\begin{aligned}
P_{k|k}=&P_{k|k-1}-\mathrm{Cov}(x_k,\varepsilon_k)\mathrm{Cov}^{-1}(\varepsilon_k)\mathrm{Cov}^T(x_k,\varepsilon_k).
\end{aligned}\end{equation}
Substituting the two equations in Lemma 2 into (69) and (70) yields (14) and (15). If $S\neq0$, it follows from Lemma 1 that
\begin{equation}\begin{aligned}
x_{k|k}=&x_{k|k-1}+\mathrm{Cov}(x_k,y_k)\mathrm{Cov}^{-1}(y_k)y_k,
\end{aligned}\end{equation}
\begin{equation}\begin{aligned}
P_{k|k}=&P_{k|k-1}-\mathrm{Cov}(x_k,y_k)\mathrm{Cov}^{-1}(y_k)\mathrm{Cov}^T(x_k,y_k).
\end{aligned}\end{equation}
Recall the definition of $y_k$, one can obtain from Lemma 2 that
\begin{equation}\begin{aligned}
\mathrm{Cov}(x_k,y_k)=P^s_{k|k-1}C_S^T,
\end{aligned}\end{equation}
\begin{equation}\begin{aligned}
\mathrm{Cov}(y_k)=C_SP^s_{k|k-1}C_S^T.
\end{aligned}\end{equation}
Substituting (73) and (74) into (71) and (72) yields (14) and (15). Note that the results above hold for any $\{\varsigma_1,\cdots,\varsigma_k\}\in\{0,1\}^k$, thus Lemma 3 also holds. The proof is completed. $\square$
\subsection{Proof of Lemma 4}
Define $K=XC^TL^T(L(CXC^T+R)L^T)^{-1}$ and $\bar{K}=KL$. Then one has
\begin{equation}\begin{aligned}
&XC^TL^T(L(CXC^T+R)L^T)^{-1}LCX\\
=&XC^TL^TK^T+KLCX-K(L(CXC^T+R)L^T)K^T\\
=&XC^T(CXC^T+R)^{-1}CX\\
&-(\bar{K}-XC^T(CXC^T+R)^{-1})(CXC^T+R)\\
&\times(\bar{K}-XC^T(CXC^T+R)^{-1})^T.
\end{aligned}\end{equation}
The proof is completed. $\square$
\subsection{Proof of Lemma 5}
Since $\bar{m}$ only has $\mu_2$ different values, we can denote it as a constant and solve the following problem
\begin{equation}\begin{aligned}
&\max_{\varsigma,S}\mathrm{Tr}((1-\varsigma)P^s_++\varsigma\bar{P}_S)\\
&\mathrm{s.t.}\ \mathrm{rank}(S)=m-a,\ 0\leq\varsigma\leq\mu_1/a
\end{aligned}\end{equation}
for different $a\in\{1,2,\cdots,\mu_2\}$. It follows from Lemma 4 that $\bar{P}_S\geq P^s_+$. Based on this, the optimal $\varsigma$ in (76) is $\mu_1/a$ because $P^s_+$ and $\bar{P}_S$ are independent of $\varsigma$. By recalling (22), it remains to solve
\begin{equation}\begin{aligned}
&\min_{S}\mathrm{Tr}(\sum^{\infty}_{i=0}A^i\Delta(S)(A^T)^i)\\
&\mathrm{s.t.}\ \mathrm{rank}(S)=m-a.
\end{aligned}\end{equation}
Since $\mathrm{Tr}(XY)=\mathrm{Tr}(XY)$ for any $X$ and $Y$ with appropriate dimensions, one can derive that
\begin{equation}
\begin{aligned}
\mathrm{Tr}(\sum^{\infty}_{i=0}A^i\Delta(S)(A^T)^i)&=\sum^{\infty}_{i=0}\mathrm{Tr}(A^i\Delta(S)(A^T)^i)\\
&=\sum^{\infty}_{i=0}\mathrm{Tr}(\Delta(S)(A^T)^iA^i)\\
&=\mathrm{Tr}(\Delta(S)W).
\end{aligned}
\end{equation}
As a result, it remains to solve
\begin{equation}\begin{aligned}
&\min_{S}\mathrm{Tr}(\Delta(S)W)\\
&\mathrm{s.t.}\ \mathrm{rank}(S)=m-a.
\end{aligned}\end{equation}
Similar to the proof of Theorem 2 in \cite{Shang9882330}, we can derive that an optimal solution to (79) is (27). The proof is completed. $\square$
\subsection{Proof of Lemma 6}
Let $w$ be an unstable eigenvalue of $A^T$, then one can deduce that $\mathrm{Re}(w^HK^s)\notin\mathrm{Ker}(w^HK^s)$ or $\mathrm{Im}(w^HK^s)\notin\mathrm{Ker}(w^HK^s)$ if $w^HK^s\neq0$. It remains to prove that $w^HK^s\neq0$. It follows from the convergence of the Kalman filter that
\begin{equation}\begin{aligned}
P^s_-=&AP^s_-A^T+BQB^T\\
&-AK^s(CP^s_-C^T+R)(K^s)^TA^T.
\end{aligned}\end{equation}
With (80), one has
\begin{equation}\begin{aligned}
&(1-|\lambda|^2)w^HP^s_-w\\
=&w^HBQB^Tw-|\lambda|^2w^HK^s(CP^s_-C^T+R)(K^s)^Tw.
\end{aligned}\end{equation}
Suppose $w^HK^s=0$. Then (81) can be simplified as
\begin{equation}\begin{aligned}
(1-|\lambda|^2)w^HP^s_-w=w^HBQB^Tw.
\end{aligned}\end{equation}
Since $|\lambda|\geq1$, $B^Tw$ is forced to be $0$. This contracts the stabilizability of $(A,B)$. The proof is completed. $\square$
\subsection{Proof of Corollary 1}
Similar to the proof of Theorem 1, one can obtain the following equation:
\begin{equation}\begin{aligned}
\mathbb{E}[P_{i,k|k}]=\tau_i\mathrm{Cov}(x_k|S_i\varepsilon_1,\cdots,S_i\varepsilon_k)+(1-\tau_i)P^s_{k|k}.
\end{aligned}\end{equation}
Then, according to the optimality condition of the Kalman filter, one has $P^s_{k|k}\leq\mathrm{Cov}(x_k|S_i\varepsilon_1,\cdots,S_i\varepsilon_k)$, which yields the first property. Moreover, when $S^T_2\in \mathrm{R}(S^T_1)$, one has $\mathcal{L}(S_2\varepsilon_1,\cdots,S_2\varepsilon_k)\subset \mathcal{L}(S_1\varepsilon_1,\cdots,S_1\varepsilon_k)$, which means
\begin{equation}\begin{aligned}
\mathrm{Cov}(x_k|S_1\varepsilon_1,\cdots,S_1\varepsilon_k)\leq\mathrm{Cov}(x_k|S_2\varepsilon_1,\cdots,S_2\varepsilon_k).
\end{aligned}\end{equation}
This gives the second property. The proof is completed. $\square$
\subsection{Proof of Lemma 7}
According to the period of $f(\cdot)$, it can be verified that  $\sum^L_{j=1}f(i+j)=|f|$ for all $i\in\mathbb{N}_+$. Then, it follows from (30) and (38) that
\begin{align}
&\frac{1}{L}\sum^{L}_{i=1}P^s_++P(S,f,i)\\
=&P^s_++\frac{|f|}{L}\sum^\infty_{j=0}A^{j}(\Delta(I)-\Delta(S))(A^T)^{j}.
\nonumber\end{align}
With (85), the optimization problem (53) is equivalent to
\begin{equation}\begin{aligned}
&\max_{\frac{|f|}{L},\bar{m},S}\frac{|f|}{L}\mathrm{Tr}(W(\Delta(I)-\Delta(S)))\\
&\mathrm{s.t.}\ \mathrm{rank}(S)=m-\bar{m},\ 0\leq\frac{|f|}{L}\bar{m}\leq\mu_3,\\
&\ \ \ \ \ \bar{m}\in\{1,2,\cdots,\mu_4\}.
\end{aligned}\end{equation}
In this case, the rest proof can be conducted similar to that of Lemma 5. The proof is completed. $\square$
\subsection{Proof of Corollary 2}
The statement 1 can be proved directly from (15). According to (85), one knows $\frac{1}{L_{\iota}}\lim_{i\rightarrow\infty}\sum^{L}_{j=1}P_{\iota,iL+j|iL+j}$ is positively correlated with $\frac{|f_{\iota}|}{L_{\iota}}$ ($\iota\in\{1,2\}$) if $\rho(A)<1$, which yields the second statement. If $S^T_2\in \mathrm{R}(S^T_1)$, then $S_2$ can be expressed as $S_2=FS_1$, where $F$ is a full-row-rank matrix. In this case, it follows from (32) and Lemma 4 that the third statement holds. The proof is completed. $\square$
\subsection{Proof of Corollary 3}
Under the stochastic intermittent strategy (10), it follows from Theorem 1 that the steady state of the covariance of the eavesdropper can be written as
\begin{equation}\begin{aligned}
\lim_{k\rightarrow\infty}\mathbb{E}[P_{k|k}]=&P^s_++\varsigma(\bar{P}_S-P^s_+)\\
=&P^s_++\varsigma\sum^\infty_{j=0}A^{j}(\Delta(I)-\Delta(S))(A^T)^{j},
\end{aligned}\end{equation}
where the second equality follows from the definitions of $P^s_+$ and $\bar{P}_S$. By comparing (85) and (87), it can be deduced that the objective values obtained in (26) and (53) are consistent if their constraints are consistent. Then, by observing (26) and (53), one knows that the constraints of (26) and (53) are the same if $\mu_1=\mu_3$ and $\mu_2=\mu_4$. The proof is completed. $\square$
\subsection{Proof of Theorem 5}
1) It follows from Lemma 3 that the covariance $P_{k|k}$ of the eavesdropper is given by
\begin{equation}\begin{aligned}
P_{k|k}=&P^s_{k|k}+A^{k-\delta}\Delta(I,\delta)(A^{k-\delta})^T,\ \forall k\geq\delta.
\end{aligned}\end{equation}
It is trivial that the sequence $\{P^s_{k|k}\}$ converges. Then, it remains to analyze the second term at the right-hand side of (88). When $\rho(A)<1$, one has $\lim_{k\rightarrow\infty}A^{k-\delta}\Delta(I,\delta)(A^{k-\delta})^T=0$, which yields the first statement.

2) Denote the Jordan decomposition of $A$ as $A=P\mathrm{diag}(J_1,J_2,\cdots,J_s)P^{-1}$, where $\rho(J_i)=|\lambda_i|$, $\lambda_i$ is the eigenvalue of $A$, $\rho(J_1)\geq\rho(J_2)\geq\cdots\geq\rho(J_s)$ and $s$ is determined by the eigenvalues and eigenvectors of $A$. Then, one has $A^{k-\delta}=P\mathrm{diag}(J_1^{k-\delta},J_2^{k-\delta},\cdots,J_s^{k-\delta})P^{-1}$.
It is clear that the stable Jordan blocks of $J^{k-\delta}$ will tend to $0$ when $k\rightarrow\infty$. Moreover, since $\rho(J_i)=\lambda_i$ if $|\lambda_i|=1$ (i.e., the unstable Jordan blocks are scalars), one can deduce that $\lim_{k\rightarrow\infty}|J_i^{k-\delta}|=1$. In this case, the sequence $\{A^{k-\delta}\}$ is bounded, which also implies that $\{P_{k|k}\}$ is bounded.

3) Let $u$ be an eigenvector of $A^T$, then one has
\begin{equation}\begin{aligned}
\mathrm{Tr}(A^{k-\delta}\Delta(I,\delta)&(A^{k-\delta})^T)\\
&\geq\frac{|\lambda|^{2(k-\delta)}}{\|u\|^2}u^T\Delta(I,\delta)u,\ \forall k\geq\delta,
\end{aligned}\end{equation}
where $A^Tu=\lambda u$. The right-hand side of (89) tends to infinity if $|\lambda|>1$ and $CP^s_{\delta|\delta-1}u\neq0$. In this case, according to PBH test, we know that a sufficient condition for unbounded $\{P_{k|k}\}$ is that $(CP^s_{\delta|\delta-1},A^T)$ is detectable. Then, it follows from the Lyapunov test for the detectability that $(CP^s_{\delta|\delta-1},A^T)$ is detectable if and only if there exists $X>0$ and $K$ such that $X\geq (A^T-KCP^s_{\delta|\delta-1})X(A^T-KCP^s_{\delta|\delta-1})^T+I$. Utilizing Schur complement lemma \cite{boyd1994linear} for this inequality yields
\begin{equation}\begin{aligned}
\begin{bmatrix}
{X} & {A^T-KCP^s_{\delta|\delta-1}} & {I}\\
{(A^T-KCP^s_{\delta|\delta-1})^T} & {X^{-1}} & {0}\\
{I} & {0} & {I}
\end{bmatrix}\geq0.
\end{aligned}\end{equation}
Then, multiplying $\mathrm{diag}(X^{-1},I,I)$ at the left and right ends of (90) respectively gives
\begin{equation}\begin{aligned}
&\begin{bmatrix}
{X^{-1}} & {X^{-1}(A^T-KCP^s_{\delta|\delta-1})} & {X^{-1}}\\
{(A^T-KCP^s_{\delta|\delta-1})^TX^{-1}} & {X^{-1}} & {0}\\
{X^{-1}} & {0} & {I}
\end{bmatrix}\\
&\geq0.
\end{aligned}\end{equation}
Finally, replacing the variables $X^{-1}$ and $X^{-1}K$ with $Y$ and $Z$ completes the proof. $\square$
\bibliographystyle{ieeetr}

\end{document}